\documentclass[letterpaper,prl,twocolumn,amsmath,amssymb,longbibliography]{revtex4-1}
\usepackage[utf8]{inputenc}
\usepackage[draft]{hyperref}
\usepackage{graphicx}% Include figure files
\usepackage{dcolumn}% Align table columns on decimal point
\usepackage{bm}
\usepackage{amssymb}
\usepackage{physics}
\usepackage{color}
\usepackage{soul}
\usepackage{epstopdf}
\usepackage{float}
\usepackage[normalem]{ulem}

\definecolor{darkgreen}{rgb}{0.0, 0.5, 0.0}

\DeclareMathAlphabet\mathbfcal{OMS}{cmsy}{b}{n}

\begin{document}
\title{Spin selection rule for {\it S} level transitions in atomic rubidium under paraxial and nonparaxial two-photon excitation}

\author{Krishnapriya Subramonian Rajasree}
\author{Ratnesh Kumar Gupta}
\author{Vandna Gokhroo}
\author{Fam~{Le~Kien}}
\author{Thomas Nieddu}
\altaffiliation[Present address: ]{Laboratoire Kastler Brossel, Sorbonne Universit\'{e}, CNRS,
ENS-Universit\'{e} PSL, Coll\`{e}ge de France, 4 place Jussieu, F-75005 Paris, France}
\author{Tridib Ray}
\altaffiliation[Present address: ]{Laboratoire Kastler Brossel, Sorbonne Universit\'{e}, CNRS,
ENS-Universit\'{e} PSL, Coll\`{e}ge de France, 4 place Jussieu, F-75005 Paris, France}
\author{S\'{i}le {Nic~Chormaic}}
\email[Corresponding author: ]{sile.nicchormaic@oist.jp}
\author{Georgiy Tkachenko}
\affiliation{Okinawa Institute of Science and Technology Graduate University, Onna, Okinawa 904-0495, Japan}

\date{\today}

\begin{abstract}
We report on an experimental test of the spin selection rule for two-photon transitions in atoms. In particular, we demonstrate that the $5S_{1/2}\to 6S_{1/2}$ transition rate in a rubidium gas follows a quadratic dependency on the helicity parameter linked to the polarization of the excitation light. For excitation via a single Gaussian beam or two counterpropagating beams in a hot vapor cell, the transition rate scales as the squared degree of linear polarization. The rate reaches zero when the light is circularly polarized. In contrast, when the excitation is realized via an evanescent field near an optical nanofiber, the two-photon transition cannot be completely extinguished (theoretically, not lower than 13\% of the maximum rate, under our experimental conditions) by only varying the polarization of the fiber-guided light. Our findings lead to a deeper understanding of the physics of multiphoton processes in atoms in strongly nonparaxial light.
\end{abstract}

\maketitle

Two-photon excitation, as proposed by Maria G\"{o}ppert-Mayer in 1931~\cite{goppertmayer_adphysik_1931} and first experimentally demonstrated in the early 1960s~\cite{kaiser_prl_1961,abella_prl_1962}, revolutionized the fields of spectroscopy~\cite{friedrich_jce_1982,schlawin_acr_2018}, fluorescence microscopy~\cite{denk_s_1990}, and optical metrology~\cite{hilico_epj_1998}. Unlike for single-photon processes, two-photon excitation gives access to energy transitions which are electric dipole forbidden. In addition, if the simultaneously absorbed photons have the same frequency and are provided by couterpropagating beams, the fluorescence spectrum is free of Doppler broadening~\cite{1970ZhPmR..12..161V,Cagnac73,biraben_prl_1974,levenson_prl_1974}; thus hyperfine transition lines can be clearly seen, allowing one to achieve a robust frequency reference~\cite{nieddu_oe_19,wang_lplett_2019}. 

Two-photon transitions in atoms can only occur between electron orbitals of the same parity. This leads to a selection rule for the allowed change of the orbital angular momentum of the electron: $\Delta L=0,\pm2$. In addition, the hyperfine energy levels involved in the transition obey a selection rule for the total angular momentum quantum number: $|\Delta F|\leq2$~\cite{salour_ap_1978}. Here, we focus on the `spin' selection rule, which further restricts the angular momentum changes in electric dipole allowed, single frequency, two-photon transitions between $S$ levels in atoms where the intermediate level is detuned from the single-photon resonance frequency. In this case, $\Delta F=0$ and $\Delta m_F=0$ (with $m_F$ being the magnetic quantum number) apply, meaning that the total spin of the atom must be preserved. If we assume that the spin of light is well-defined, conservation of  angular momentum in the excitation process requires that the two  photons have mutually cancelling spin projections along the quantization~axis.

This principle has been verified experimentally using sodium~\cite{biraben_prl_1974,levenson_prl_1974} or rubidium~\cite{nieddu_oe_19,wang_lplett_2019} vapor  illuminated by counterpropagating Gaussian beams. Doppler-free transition peaks were observed when the beams had equal linear polarizations or opposite circular polarizations in the laboratory frame, but the peaks disappeared for circular polarizations of the same handedness. When the excitation light is elliptically polarized, two-photon transitions are not completely absent, but occur at a rate that depends on the shape of the polarization ellipse, as demonstrated experimentally~\cite{wang_lplett_2019}.

It is important to note that, in the forementioned spectroscopy experiments, the atoms interacted with paraxial, free-space electromagnetic fields. In recent years, there has been significant  interest in shifting toward light fields confined at the micro- or even nanoscale, and, in one of the more popular systems,
neutral atoms are coupled to the evanescent field of a vacuum-clad optical nanofiber (ONF) ~\cite{Mitsch2014,Ruddell:17,PhysRevA.96.043859, PhysRevX.8.031010,nayak_prl_2019,white_prl_2019,Corzo2019, jones_prl_2020}. Strong confinement of light around the ultrathin ONF waist region has led to demonstrations of nonlinear atomic processes at sub-$\mu$W excitation powers~\cite{PhysRevA.92.043806,kumar_pra_2015,kumar_njp_2015} and an electric quadrupole transition driven by a few~$\mu$W~\cite{ray_njp_2020}.
Owing to the recent achievements of full polarization control for light guided in single-mode nanofibers~\cite{lei_prappl_2019,joos_oe_2019,tkachenko_jo_2019}, the spin selection rule for two-photon excitation in  ONF-coupled atoms can now be explored experimentally. 

In this work, we first develop a theoretical model to describe the spin selection rule for two-photon transitions as a function of the polarization of the excitation light.  Furthermore, as a verification tool, we experimentally study the polarization dependence of an $S\to S$ two-photon transition in a $^{87}$Rb gas for two conceptually different excitation conditions: (i)~warm atoms in a vapor cell with Gaussian beam illumination, and (ii)~laser-cooled atoms in the evanescent field of a single-mode ONF, where the light is strongly nonparaxial. 

We consider an atomic transition from a lower state $\ket{g}$ to an upper state $\ket{e}$ (with corresponding frequency, $\omega_{eg} = \omega_e-\omega_g$) excited by the simultaneous absorption of photons from two light fields
%labeled by the index $j=1,2$ and 
characterized by a frequency, $\omega_j$, amplitude, ${\mathbfcal{E}}_j$, and a unit polarization vector,~${\bf u}_j$. Here, we discuss only the case when the two fields are identical ($\omega_1=\omega_2=\omega_{eg}/2$; ${\mathcal{E}}_1={\mathcal{E}}_2={\mathcal{E}}/\sqrt{2}$; ${\bf u}_1 = {\bf u}_2 = {\bf u}$), up to the direction of propagation with respect to the $z$ axis of the Cartesian coordinate system, $(x,y,z)$ (see \cite{supplementary} for more details). When the two fields are counterpropagating, we denote the configuration as `2-beam' and `1-beam' otherwise. For such a single frequency, two-photon excitation, the transition rate is
\begin{equation}
     P_{ge}\propto \frac{1}{\Gamma}|\alpha_J^{(0)}|^2\xi,
     \label{rate}
\end{equation}
where $\Gamma$ is the total decay rate, $\alpha_J^{(0)}$ is the reduced scalar coefficient \cite{supplementary} and the factor $\xi=|({\mathbfcal{E}}\cdot{\mathbfcal{E}})|^2 = |{\cal E}|^4 |({\bf u}\cdot{\bf u})|^2$ expresses both of the characteristic features of the two-photon process: the quadratic dependency on the field intensity, $I = |{\cal E}|^2$, and the polarization dependency. In this work, we focus on the latter.

The excitation field can be written as a function of the helicity parameter, $-1\leq\sigma\leq1$~\cite{bliokh_prl_2008,tkachenko_optica_2020}, as follows:
\begin{equation}
    {\mathbfcal{E}} = \frac{1}{\sqrt{2}}\left(\sqrt{1 + \sigma} {\mathbfcal{E}}_{+1} +\sqrt{1 - \sigma} {\mathbfcal{E}}_{-1}\right),
\end{equation}
where ${\mathbfcal{E}}_p=(e_r\hat{r}+p e_{\varphi}\hat\varphi+e_{z}\hat{z})e^{i(p\varphi+\beta z)}$, $p=\pm1$ is the polarization index for quasi-circularly polarized fundamental guided modes~\cite{le_kien_pra_2013}, $\beta$ is the propagation constant, and $e_r$, $e_\varphi$, $e_z$ are the reduced cylindrical components of the mode function and they are independent of $\varphi$ and the $z$ direction~\cite{lekien_oc_2004,tong_oe_2004,lekien_pra_2017}.
Note that, if the field is retro-reflected, one must change the sign for $\beta$, $\sigma$, and $p$.

After statistical averaging of $\xi$ over the atomic position, we find \cite{supplementary}

\begin{equation}
P_{ge}\propto\bar{\xi}= A-\sigma ^2 B,
\label{eq:xibar}
\end{equation}

\noindent where
\begin{align}
    A = &{\expval{(|e_r|^2+|e_{\varphi}|^2-|e_z|^2)^2}}_r\nonumber\\&+0.5\,{\expval{(|e_r|^2-|e_{\varphi}|^2-|e_z|^2)^2}}_r\,,\nonumber\\
    B = &{\expval{(|e_r|^2+|e_{\varphi}|^2-|e_z|^2)^2}}_r\nonumber\\&-0.5\,{\expval{(|e_r|^2-|e_{\varphi}|^2-|e_z|^2)^2}}_r\,,
    \label{eq:AB}
\end{align}
and $\expval{...}_r$ stands for statistical averaging over the radial distance $r$ from the nanofiber. The quadratic Eq.~\ref{eq:xibar} predicts that the transition rate is maximum for linearly polarized ($\sigma=0$) and minimum (but, in general, {\it nonzero}) for circularly polarized ($\sigma=\pm1$) excitation.

%------------------------------------------------
\begin{figure}
\centering
\includegraphics[width=1\linewidth]{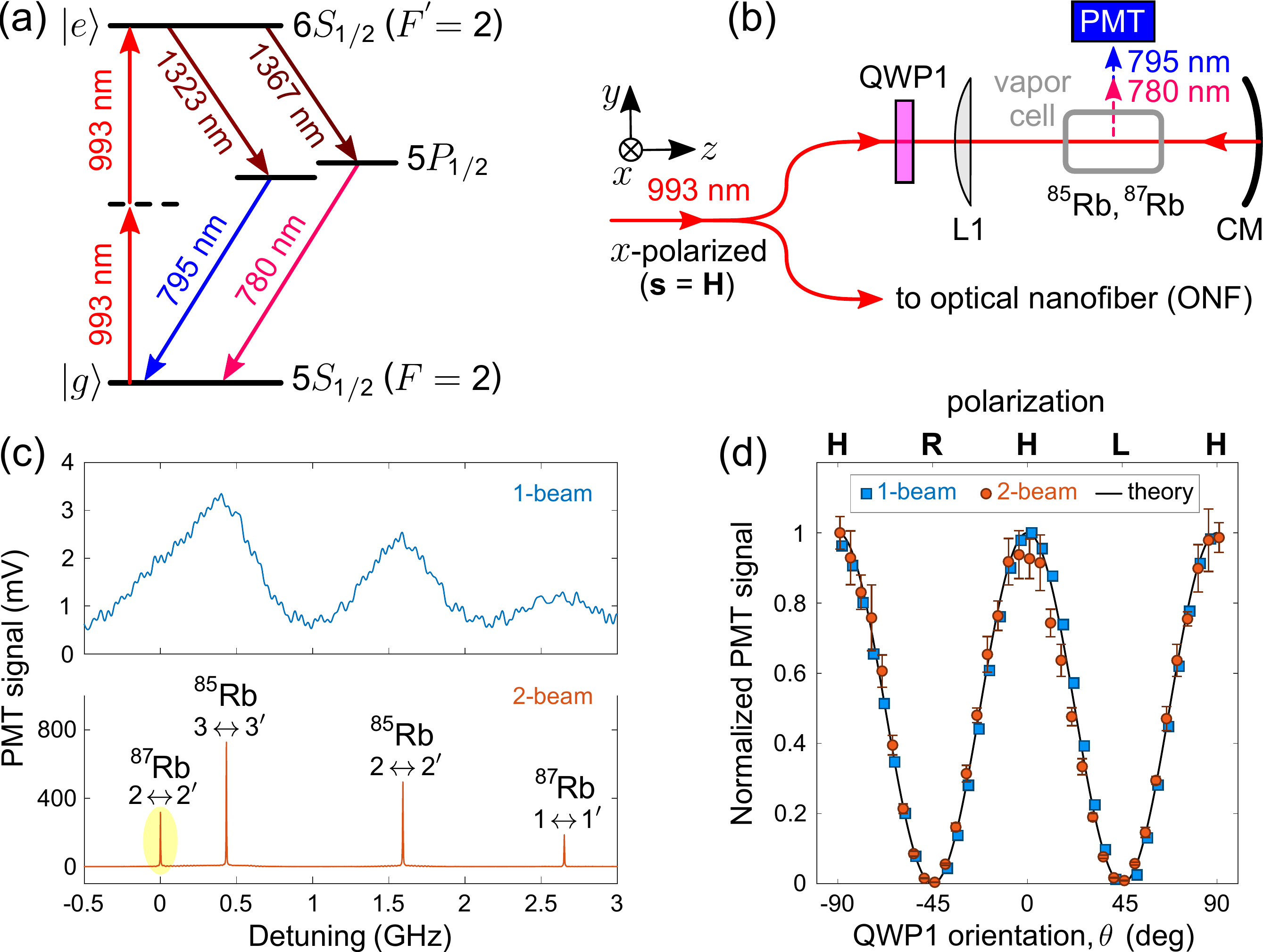}
\caption{(a)~Simplified energy level diagram for $^{87}$Rb as relevant to the experiment. (b)~Vapor cell spectroscopy setup. QWP1: quarter-wave plate, L1: convex lens, CM: concave mirror, PMT: photomultiplier tube. (c)~Spectroscopy signal (background subtracted) measured in the 1-beam~(top) and 2-beam~(bottom) configurations. Labeled peaks correspond to the Doppler-free hyperfine transitions, with the chosen one ($2\leftrightarrow2'$ in $^{87}$Rb) highlighted. (d)~Polarization dependence of the fluorescence signal from the chosen transition. Squares and circles: experimental data for 1-beam and 2-beam configurations, respectively. Solid line: simulation. Indicated principal polarization states of the 993~nm light: horizontal ($\bf H$), left- and right-handed circular ($\bf L$, $\bf R$). }
\label{fig:vaporcell}
\end{figure}
%------------------------------------------------

In order to verify the above theoretical result experimentally, we chose the  $5S_{1/2}(F=2)\to6S_{1/2}(F^{'}=2)$ transition  in $^{87}$Rb, see Fig.~\ref{fig:vaporcell}(a), accessible via two-photon excitation at 993~nm \cite{nieddu_oe_19}. The excitation light is provided by a Ti:Sapphire laser and its frequency is fine-tuned via spectroscopy in a vapor cell containing a natural mixture of $^{85}$Rb and $^{87}$Rb atoms, maintained at 100$^\circ$C. A schematic of the experiment is given in Fig.~\ref{fig:vaporcell}(b). The polarization of the excitation beam is given by a unit vector, ${\bf s} = (1,S_1,S_2,S_3)$, where $S_{1,2,3}$ are the reduced Stokes parameters defined from the point of view of the receiver. A  993~nm laser beam of 150~mW is weakly focused into the vapour cell by a convex lens (L1, focal distance $f_1=150$~mm) and the 795~nm and the 780~nm fluorescence (from the $5P_{1/2}\to5S_{1/2}$ and the $5P_{3/2}\to 5S_{1/2}$ decay paths, respectively) emitted at the focal point is detected by means of a photomultiplier tube (PMT) through a relay telescope and a shortpass filter (FES0800 from Thorlabs, not shown). The fluorescence intensity is a measure of the two-photon transition rate. Figure~\ref{fig:vaporcell}(c) (upper panel) depicts the spectroscopy signal collected for horizontal polarization, ${\bf H}=(1,1,0,0)$ (note that a low-pass digital frequency filter was applied to the data in order to suppress the noise). To convert this 1-beam configuration into the 2-beam case, we add a concave mirror (CM, $f_{\rm CM} = f_1/2=75$~mm) placed at $2f_1$ away from~L1. As a result, the spectrum reveals the 200-fold increased Doppler-free peaks, see the lower panel in Fig.~\ref{fig:vaporcell}(c), where each hyperfine transition of $^{85}$Rb and $^{87}$Rb is indicated with the corresponding $F\leftrightarrow F'$ values.

Fluorescence signals from the vapor cell for both the 1-beam and 2-beam configurations should follow~Eq.~\ref{eq:xibar}, irrespective of the Doppler broadening in the former case. To test this, we introduce a quarter-wave plate, QWP1, before L1 and scan the helicity parameter, $\sigma$, over the whole $[-1,1]$ range by varying $\theta$, the angle between $x$ and the slow axis of the wave plate. As a result, the initial horizontal polarization state is transformed into $(1,\cos^2{2\theta},\sin{2\theta}\cos{2\theta},-\sin{2\theta})$ and $\sigma = -S_3 = \sin{2\theta}$. Since the Gaussian field is paraxial, we assume $e_z=0$ and $|e_r|=|e_{\varphi}|$.  Thence,  Eq.~\ref{eq:AB} gives $A=B$ and Eq.~\ref{eq:xibar} reduces to $P_{ge}\propto 1-\sigma^2$. The two-photon transition rate is expected to scale as $\cos{^2 2\theta}$, reaching zero at $\theta=(2n+1)\pi/4$ with $n\in \mathbb{Z}$, as confirmed by the measured polarization dependence of the fluorescence for the chosen two-photon transition, see Fig.~\ref{fig:vaporcell}(d). In the 1-beam case (blue squares), the signal is defined simply as the mean voltage output of the PMT and the confidence range is smaller than the marker size. In the 2-beam configuration, the transition spectral profile is fitted to a Lorentzian curve~\cite{biraben_jpf_1979} and the markers (orange circles) shown in Fig.~\ref{fig:vaporcell}(d) are the average maxima of the fitted curves, while the error bars indicate the standard deviation range over 10 independent measurements. For convenience of presentation, each data set in Fig.~\ref{fig:vaporcell}(d) is normalized to the maximum.

It is important to note that the results obtained with the hot vapor cell can be explained in simpler terms. Since the transition is allowed only when the excitation light carries zero net spin angular momentum, $P_{ge}$ is intuitively expected to be linked to the degree of linear polarization, ${\rm DOLP} = \sqrt{S_1^2+S_2^2}$. Physically, the DOLP is the maximum fraction of optical power one can measure in transmission through a lossless linear polarizer. Owing to the quadratic power dependence of the two-photon transition rate, $P_{ge}\propto {\rm DOLP}^2 = \cos^4{2\theta}+\sin{^2 2\theta}\cos{^2 2\theta} = \cos{^2 2\theta}$, which we indeed confirm experimentally.
Continuing this line of thought, we notice that the mentioned lossless linear polarizer defines the DOLP for both the forward-propagating beam and the retro-reflected one. Thus, if the beams are linearly polarized in planes titled by an angle $\delta$ with respect to each other, the two-photon transition rate would scale as ${\rm DOLP}^2=\cos^2 \delta$, which is equivalent to Malus's law.
%, albeit for such two-photon transitions excited via counterpropagating beams.
This result explains the absence of Doppler-free peaks in the crossed polarization ($\pi$-$\pi'$) case experimentally tested in~\cite{nieddu_oe_19}, where the setup contained a quarter-wave plate between the vapor cell and the concave mirror.

%------------------------------------------------
\begin{figure}
\centering
\includegraphics[width=1\linewidth]{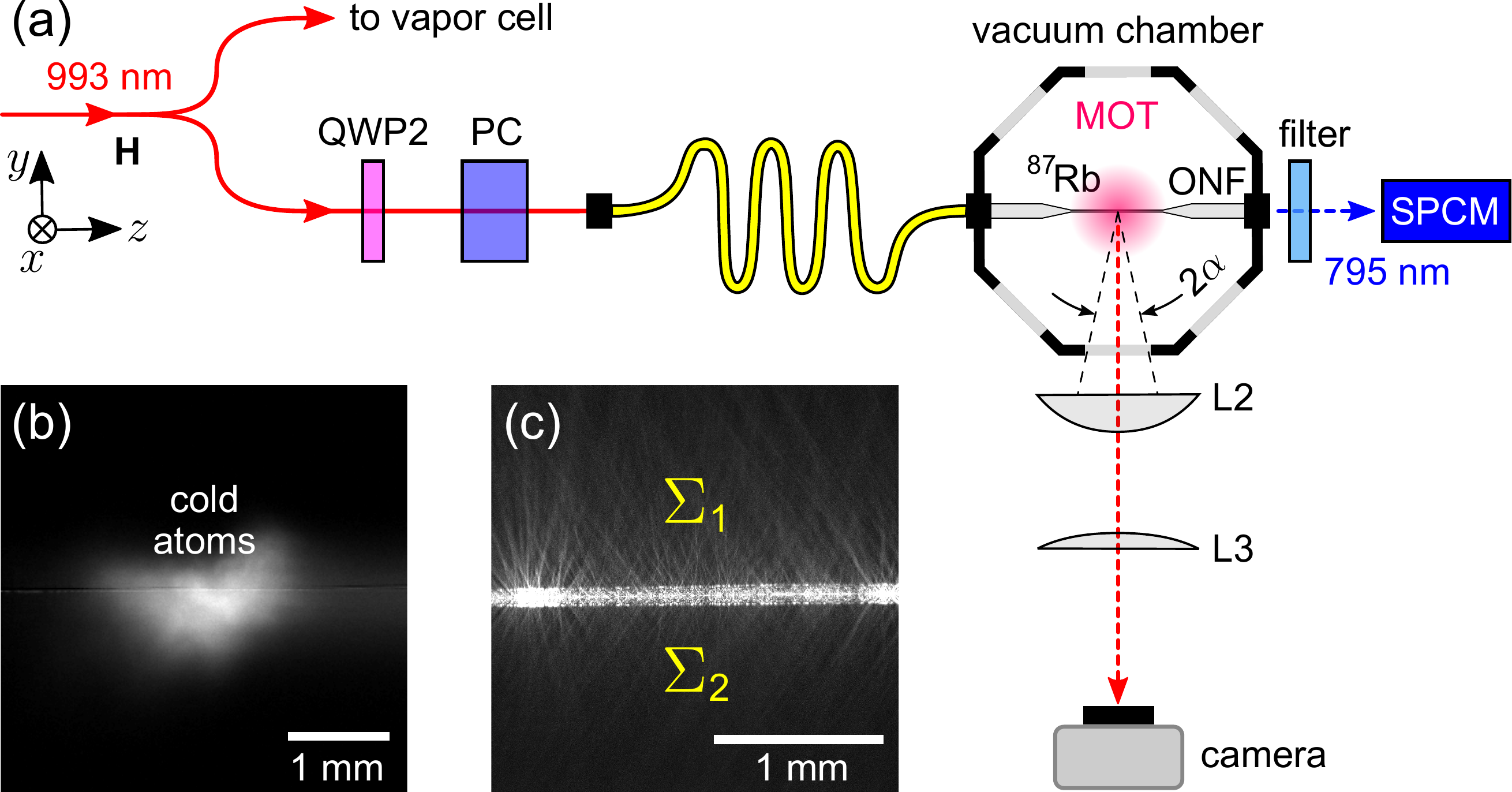}
\caption{(a)~Atom-nanofiber experimental setup. QWP2: quarter-wave plate, PC: polarization compensator, L2, L3: convex lenses, SPCM: single photon counting module  (b)~Fluorescence image of the cold atom cloud around the nanofiber. (c)~Typical image of the excitation light scattered at the nanofiber waist. $\Sigma_1$ and $\Sigma_2$ are the brightness sums of the upper and lower halves of the image.}
\label{fig:nanofiber}
\end{figure}
%------------------------------------------------

Now, let us consider the case where excitation is by the evanescent field of light guided in an optical nanofiber. The experimental setup is sketched in Fig.~\ref{fig:nanofiber}(a). The ONF is fabricated from a commercial, single-mode optical fiber (SM800-5.6-125, Fibercore) via a flame-brushing technique~\cite{ward_rsinstru_2014}. The initial fiber diameter of $125\,\mu$m is exponentially tapered to 400~nm at the waist, thereby supporting only the fundamental guided modes for both the 993~nm and 795~nm wavelengths (the transmission at 993~nm is about 30\%). The ONF is mounted in an ultrahigh vacuum chamber and a standard 3-beam retro-reflected magneto-optical trap (MOT) is used to produce the cold $^{87}$Rb atoms at the nanofiber waist, see~\cite{rajasree_prrea_2020} for details. An average atom cloud density of 10$^{9}$/cm$^3$ is estimated from fluorescence imaging of the cold atoms, see Fig.~\ref{fig:nanofiber}(b), using a CCD camera (Andor Luca$^{\rm EM}$R DL-604M-OEM) and a microscope composed of two convex lenses, L2 ($f_2=125$~mm) and L3 ($f_3=250$~mm). The 993~nm light is sent through the fiber for two hours prior to  measurements in order to heat the ONF and thus prevent deposition of atoms onto its surface. The optical power, measured at the output end of the fiber, is maintained at 0.6~mW during the experiment. In a typical experiment sequence, the MOT is loaded to saturation, then the excitation laser is scanned around the chosen transition, while the 795~nm fluorescence  is coupled into the nanofiber and recorded by a single-photon counting module (SPCM) through a band-pass filter. In a similar manner to the vapor cell tests, in this setup the input polarization is varied by means of a quarter-wave plate, QWP2, placed before the fiber coupler. In a steady state (i.~e., after the initial 2-hour heating stage, and when QWP2 is not rotating), the polarization is stable within $1^{\circ}$ on the Poincar{\'e} sphere, as confirmed by a free-space polarimeter (PAX1000IR, Thorlabs) placed at the output pigtail of the fiber.

Due to stress-induced birefringence in the tapered fiber, the preset input polarization state of the 993~nm fiber-guided light is transformed into an unknown state at the nanofiber waist. In order to gain control over the helicity at the waist, we reverse the transformation using a polarization compensator, PC, introduced between QWP2 and the input pigtail of the fiber. The compensator consists of a variable retarder (liquid crystal type, slow axis parallel to~$x$) and two quarter-wave plates~\cite{lei_prappl_2019}. The compensation procedure is based on imaging  the 993~nm light (through an $x$-oriented linear polarizer, not shown) scattered from natural imperfections of the nanofiber~\cite{tkachenko_jo_2019}. This method relies on identifying two non-orthogonal linear polarization states at the nanofiber waist, the first being horizontal (or vertical) and the second being diagonal (or anti-diagonal). The latter corresponds to the absolute maximum (or minimum) of the intensity difference between two regions of the scattered light image, $\Delta = \Sigma_1-\Sigma_2$, see Fig.~\ref{fig:nanofiber}(c). To maximize the precision of $\Delta$ measurements, we replace L2 by a convex lens with a diameter of 50~mm and a focal distance of 75~mm, thus achieving a maximum collection angle, $\alpha\approx18^{\circ}$. The errors in the state identification are expected to be less than~$10^{\circ}$ on the Poincar{\'e} sphere~\cite{tkachenko_jo_2019}. When $|\sigma|$ approaches unity (the most interesting case in this study), this error corresponds to a confidence range of about $1.5$\% for $|\sigma|$ and $3$\% for $\sigma^2$.

%------------------------------------------------
\begin{figure}
\centering
\includegraphics[width=1\linewidth]{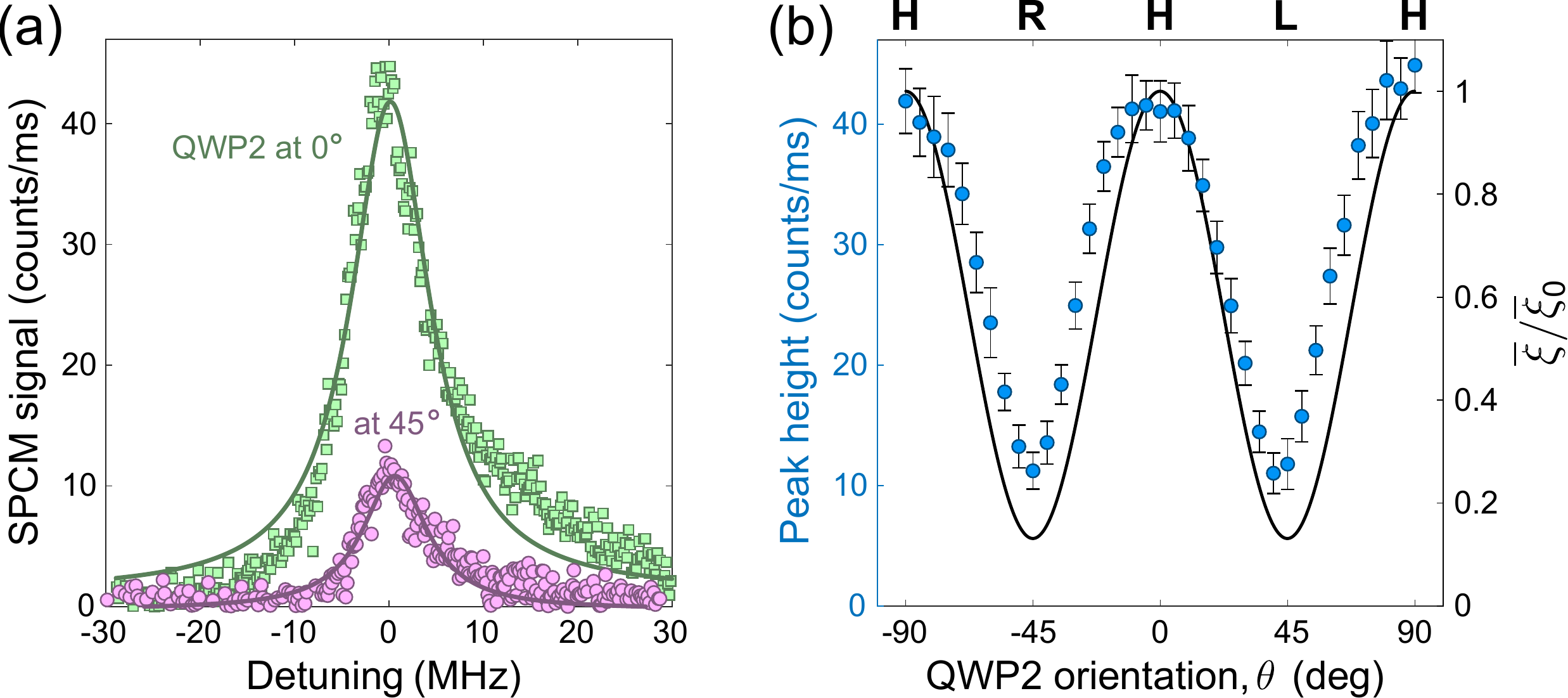}
\caption{Measured 795~nm fluorescence from the cold atoms. (a)~Typical transition peaks for a linearly polarized (QWP2 at $\theta=0^{\circ}$, $\sigma=0$) and a circularly polarized ($\theta=45^{\circ}$, $\sigma=1$) input 993~nm beam. The solid lines represent Lorentzian curve fitting. (b)~Polarization dependence of the 795 nm fluorescence. Solid line: simulation.}
\label{fig:results}
\end{figure}
%------------------------------------------------

The experimental results are shown in Fig.~\ref{fig:results} where panel~(a) depicts the 795~nm fluorescence collected by the SPCM in the two limiting cases: the excitation light is (i) linearly polarized ($\sigma=0$, ${\bar \xi}_0=A$) and (ii) circularly polarized ($\sigma=\pm1$, ${\bar \xi}_{\pm1}=A-B$). Each set of data is fitted to a Lorentzian curve (solid lines in Fig.~\ref{fig:results}(a)), and its peak value is the measure of the two-photon transition rate.~Since guided modes of a nanofiber feature a nonvanishing longitudinal field component, $A\neq B$ and ${\bar \xi}\neq 0$, even when the input light is circularly polarized. The fluorescence measurements over the complete range of $\sigma$ are summarized in Fig.~\ref{fig:results}(b) where the solid curve is the calculated ${\bar \xi}/{\bar \xi}_0$ and the error bars indicate the standard deviation ranges obtained from 40~experimental sequences for every orientation of QWP2.

We attribute the discrepancy between the experimental data and the theoretical curve, specifically the shallower and narrower dips in the measurement, to several experimental factors beyond our control: the polarization compensation only works for the transverse field components, atoms are not necessarily evenly distributed around the nanofiber (as suggested by the cloud picture in Fig.~\ref{fig:nanofiber}(b)), individual atoms may see local inhomogeneities of the excitation field near the ONF waist, or residual magnetic field and cooling beams may be present in the excitation region, thereby influencing the two-photon process.
The lateral shift of the rising slopes seen in both periods of the $\theta$ dependence in Fig.~\ref{fig:results}(b) is likely to be an experimental artefact such as imperfection of the waveplate, enhanced by coupling of light into the fiber.
Other effects not taken into account are possible polarization-dependent saturation~\cite{spillane_2008} of the transition in the atomic cloud, the effect of van der Waals' forces near the fiber surface~\cite{Frawley_2012,Kien_2018}, a polarization-induced inhomogeneity in the intensity profile~\cite{lekien_oc_2004} and the related change in the local atomic density due to the dipole force, and position-dependent Stark shifts in the atomic energy levels~\cite{metcalf99}.  We also note that the relation $\sigma = \sin (2\theta)$ may not be exactly fulfilled for ONF-mediated excitation. For instance, the generation of orbital angular momentum in the evanescent field, which is more significant for quasi-circular polarization~\cite{le_kien_pra_2006,tkachenko_optica_2020}, effectively changes the helicity and its relation to the polarization of light sent into the fiber. This invites further studies on two-photon processes under nonparaxial fields, inclusive of the orbital degree of freedom.\\
\indent In conclusion, we have observed the spin selection rule as applied to an $S \to S$ two-photon  atomic transition, both within and beyond the paraxial limit for the excitation light. In the latter case, the light was delivered by the evanescent field of a single-mode optical nanofiber, which also served as a detection channel for the fluorescence signal, a measure of the two-photon transition rate. Owing to the accurate  polarization control at the nanofiber waist, we were able to study the transition rate as a function of helicity of the excitation. In contrast to the paraxial case, the two-photon transition in the evanescent field could not be extinguished by simply varying the polarization of the coupled light; we observed a minimum rate of about 13\% of the maximum in theory and 25\% in practice. These findings are important in the context of quantum technologies progressing towards integrated photo-emitting networks where strongly confined fields and nonlinear processes are exceedingly more common.

\indent We thank Antoine Pichen{\'e} and Ren{\'e} Henke for assembling the hot vapor cell setup. This work was financially supported by the Okinawa Institute of Science and Technology Graduate University and the Japan Society for the Promotion of Science (JSPS) Grant-in-Aid for Scientific Research (C) Grant Number 19K05316. G.~T.~was supported by the JSPS as an International Research Fellow (Standard, ID~No.P18367).

\bibliography{biblio}
\end{document}